\documentclass[12pt]{iopart}
\usepackage[english]{babel}
\usepackage[utf8]{inputenc}
\usepackage{graphicx,epstopdf}
\usepackage[colorinlistoftodos]{todonotes}
\usepackage{epsfig,color,textcase}
\usepackage{hyperref}
\usepackage{doi}
\begin{document}

\title[Hellmann–Feynman forces within Wannier functions basis]{Hellmann–Feynman Forces within the DFT+U in Wannier functions basis}
\author{D. Novoselov$^{1}$, Dm.M. Korotin$^{1}$, V.I. Anisimov$^{1,2}$}

\address{$^1$ Institute of Metal Physics, S.Kovalevskoy St. 18, 620137 Yekaterinburg, Russia}
\address{$^2$  Ural Federal University, Mira St. 19, 620002 Yekaterinburg, Russia}
\ead{novoselov@imp.uran.ru}
\begin{abstract}
The most general way to describe localized atomic-like electronic states in strongly correlated compounds is to use Wannier functions. 
In the present paper we continue the development of widely-used DFT+U method onto Wannier function basis set and propose a technique to calculate a Hubbard contribution to an atomic forces. 
The technique was implemented as a part of plane-waves pseudopotential code Quantum-ESPRESSO and tested on a charge transfer insulator NiO. 
\end{abstract}


\section{Introduction}

Quantitative description of micro- and macroscopic properties for materials with strong electron-electron interactions is a challenge for condensed matter physics during few last decades. The Density Functional Theory extensions such as DFT+U and DFT+DMFT, that allow to take into account electronic correlations, are under intensive development at the moment. Core of the DFT+correlations methods is a merging of the density functional and a model approaches. Hubbard model for correlated states is build on a result of {\em ab-initio} DFT calculation. Some localized atomic wavefunctions are chosen to describe correlated electrons. Then various Hubbard corrections for potential, charge density, total energy, atomic forces, {\em etc.} are added to the DFT calculation results. An important point in this approach is a choice of the localized atomic wavefunctions.

Many researchers used Wannier function basis for strongly correlated materials calculations in recent years, for example, see Refs.~\cite{Leonov2008,Pavarini2004,Yin2006,Lechermann2006}. 
This choice is convenient to describe a correlation effects in compounds with mixed character partially filled energy bands due to the fact that it is localized in the direct space and is a superposition of atomic orbitals of neighboring atoms.
In our previous works~\cite{Korotin2012,Korotin2013} we have formulated the DFT+U approach with Wannier functions (WF) for correlated states description. In the present paper we extend the approach with an atomic forces calculation technique.
This will allow to evaluate a phonon spectra and perform a molecular dynamics simulations for a correlated materials with complex energy bands structure near a Fermi level.

The presented technique is formulated for pseudopotential plane-wave method and was implemented as a part of Quantum-ESPRESSO package~\cite{Giannozzi2009}. We have successfully tested the approach on a typical correlated system - nickel oxide.

\section{Method}

The Hubbard contribution to the total energy could be expressed \cite{PhysRevB.71.035105} as:
\begin{eqnarray}
E_{U}[\left\{  n^{I\sigma}_{mn} \right\}] = \frac{U}{2}\sum_{Im\sigma}\Bigl( n^{I\sigma}_{mm} - \sum_{n} n^{I\sigma}_{mn} n^{I\sigma}_{nm} \Bigl),
\end{eqnarray}
where $U$ is the on-site Coulomb interaction parameter and $n^{I\sigma}_{mn}$ is the correlated states occupation matrix for site $I$ and spin $\sigma$.
By using Hellmann-Feynman theorem one can write a Hubbard contribution to an atomic forces as~\cite{PhysRev.56.340}:
\begin{equation}
\label{fundamental}
F^{U}_{\alpha i} = -\frac{\partial E_{U}}{\partial \tau_{\alpha i}} = 
- \sum_{Imn\sigma} \frac{\partial E_{U}}{\partial n^{I\sigma}_{mn}} \frac{\partial n^{I\sigma}_{mn}}{\partial \tau_{\alpha i}},
\end{equation}
where $\delta\tau_{\alpha i}$ is the displacement of the atom $\alpha$ of the unit cell in the $i$-th direction.
Therefore the Hubbard contribution to the atomic forces is:
\begin{equation}
\label{Hubforce}
F^{U}_{\alpha i} = - \frac{U}{2} \sum_{Imn\sigma} (\delta_{mn} -2n^{I\sigma}_{nm}) \frac{\partial n^{I\sigma}_{mn}}{\partial \tau_{\alpha i}}.
\end{equation}
The key point in the atomic forces calculation is an accurate computation of the occupation matrix derivative. The occupation number operator for $\mathbf{k}$-point in reciprocal space is:
\begin{equation}
\hat n^{\sigma}_{\mathbf{k}} = \sum_{\nu}\theta(\varepsilon_{\nu}-E_{f})|\psi^{\sigma}_{\mathbf{k}\nu}\rangle\langle\psi^{\sigma}_{\mathbf{k}\nu}|,
\end{equation}
where $\epsilon_\nu$ is the $\nu$-th band energy, $|\psi^{\sigma}_{\mathbf{k}\nu}\rangle$ is $\nu$-th eigenvector of the Hamiltonian matrix and $E_f$ is the Fermi energy.
The occupation matrix element in WFs basis is ({\bf k}-point weight is omitted here and below for simplicity):
\begin{equation}
\label{occup}
n^{I\sigma}_{mn}=\sum_{\mathbf{k}}\langle W^{I}_{m\mathbf{k}}|\hat n^{\sigma}_{\mathbf{k}}|W^{I}_{n\mathbf{k}}\rangle
= \sum_{\mathbf{k}\nu}\theta(\varepsilon_{\nu}-E_{f})\langle W_{m\mathbf{k}}^{I} | \hat S | \psi^{\sigma}_{\mathbf{k}\nu}\rangle\langle\psi^{\sigma}_{\mathbf{k}\nu}| \hat S | W_{n\mathbf{k}}^{I} \rangle.
\end{equation}
WFs used here are defined as a projections of Bloch sums of the atomic orbitals $|\phi^{I}_{m\mathbf{k}} \rangle$ onto a subspace of the Bloch functions (the detailed description of WFs construction procedure within pseudopotential method is given in Ref.~\cite{Korotin2008}):
\begin{equation} 
|\tilde W^{I}_{m\mathbf{k}}\rangle = \sum_{\mu=N_1}^{N_2}|\psi^{\sigma}_{\mathbf{k}\mu}\rangle\langle\psi^{\sigma}_{\mathbf{k}\mu}|\hat S|\phi^{I}_{m\mathbf{k}}\rangle,
\end{equation}
where $\hat S$ is an overlap operator of the ultrasoft pseudopotential formalism \cite{Vanderbilt1986} and it has the following form:
\begin{equation}
\hat S = \hat 1 + \sum_{stI} | \beta^{I}_{s}\rangle q^{I}_{st} \langle\beta^{I}_{t}|,
\end{equation}
where $| \beta^{I}_{s}\rangle $ is a projector function that satisfies the condition
$\langle\beta^{I}_{s} | \tilde \phi^{I}_{t}\rangle_{r<r_c} = \delta_{st}$
and
$q^{I}_{st} = \langle \phi ^{I}_{s\mathbf{k}} | \phi ^{I}_{t\mathbf{k}} \rangle - 
\langle \tilde \phi ^{I}_{s\mathbf{k}} | \tilde \phi ^{I}_{t\mathbf{k}} \rangle$
within some cut-off atomic radius (${r<r_c}$).
The $\tilde \phi ^{I}_{t\mathbf{k}}$ symbol correspond a pseudo atomic wave function inside the sphere with a radius equals $r_c$; $|\phi ^{I}_{t\mathbf{k}}\rangle=|\tilde \phi ^{I}_{t\mathbf{k}}\rangle, \forall r: r<r_c$.
Orthonormality condition in ultrasoft pseudopotential formalism has the following form:
$\langle\psi^{\sigma}_{\mathbf{k}\mu}|\hat S|\psi^{\sigma}_{\mathbf{k}\nu}\rangle = \delta_{\mu\nu}$. Values of $\beta^{I}_{s}$ and $q^{I}_{st}$ are defined within a pseudopotential generation for selected atom and this quantities aren't changed during self-consistent calculation\footnote{Detailed description of the ultrasoft pseudopotential generation procedure can be found in \cite{Martin2004}}.

We force WF to be orthonormal with standard procedure:
\begin{eqnarray} 
|W^{I}_{m\mathbf{k}}\rangle = \sum_{m'} O^{-\frac{1}{2}}_{mm'}|\tilde W^{I}_{m'\mathbf{k}}\rangle,\nonumber\\
O_{mm'} = \langle\tilde W^{I}_{m\mathbf{k}}|\tilde W^{I}_{m'\mathbf{k}}\rangle.
\end{eqnarray}
The WF's overlap matrix $O_{mm'}$ could be written as
\begin{equation}
O_{mm'} = \sum_{\mu=N_1}^{N_2}\langle\phi^{I}_{m\mathbf{k}}|\hat S|\psi^{\sigma}_{\mathbf{k}\mu}\rangle\langle\psi^{\sigma}_{\mathbf{k}\mu}|\hat S|\phi^{I}_{m'\mathbf{k}}\rangle.
\end{equation}
Lets introduce a notation $P^{I\sigma}_{m\mu}$ for the matrix:
\begin{equation}
\label{projection}
P^{I\sigma}_{m\mu}\equiv\langle\phi^{I}_{m\mathbf{k}}|\hat S|\psi^{\sigma}_{\mathbf{k}\mu}\rangle.
\end{equation}
Then the WF's overlap matrix is:
\begin{eqnarray}
O_{mm'} = \sum_{\mu=N_1}^{N_2}P^{I\sigma}_{m\mu}(P^{I\sigma}_{m'\mu})^{*},
\end{eqnarray}
and occupation matrix (\ref{occup}) can be written as:
\begin{eqnarray}
n^{I\sigma}_{mn}= \sum_{\mathbf{k}}\sum_{\mu=N_1}^{N_2}\theta(\varepsilon_{\mu}-E_{f})\sum_{m'}
(O^{-\frac{1}{2}}_{mm'})^{*}
P^{I\sigma}_{m'\mu}\sum_{n'}(P^{I\sigma}_{n'\mu})^{*}O^{-\frac{1}{2}}_{nn'}.
\end{eqnarray}
The derivative of the occupation matrix on atomic displacements takes the form:
\begin{eqnarray}
\label{dntau}
\frac{\partial n^{I\sigma}_{mn}}{\partial \tau_{\alpha i}} & = & 
\sum_{\mathbf{k}}\sum_{\mu=N_1}^{N_2}\theta(\varepsilon_{\mu}-E_{f})
\Biggl(
\sum_{m'}(O^{-\frac{1}{2}}_{mm'})^{*}\frac{\partial P^{I\sigma}_{m'\mu}}{\partial \tau_{\alpha i}}
\sum_{n'}(P^{I\sigma}_{n'\mu})^{*}O^{-\frac{1}{2}}_{nn'}+\nonumber\\
& + & \sum_{m'}(O^{-\frac{1}{2}}_{mm'})^{*}P^{I\sigma}_{m'\mu}
\sum_{n'}\Bigl(\frac{\partial P^{I\sigma}_{n'\mu}}{\partial \tau_{\alpha i}}\Bigr)^{*}O^{-\frac{1}{2}}_{nn'}
\Biggr).
\end{eqnarray}

In general case it is difficult to find accurate analytical solution for the derivative $\partial O^{-\frac{1}{2}}_{mm'} / \partial \tau_{\alpha i}$. Therefore we have neglected corresponding terms and they are not included in equation~(\ref{dntau}).
Nevertheless, if necessary it is possible to evaluate this value approximately by assuming that the overlap matrix $O$ has diagonal form.
This assumption allows one to write:
\begin{equation}
\label{dO12tau}
\frac{\partial O^{-\frac{1}{2}}_{mm}}{\partial \tau_{\alpha i}}=
-\frac{1}{2} \frac{\partial O_{mm}}{\partial \tau_{\alpha i}}O^{-\frac{3}{2}}_{mm},
\end{equation}
where
\begin{eqnarray}
\label{dOtau}
\frac{\partial O_{mm}}{\partial \tau_{\alpha i}} = 
\sum_{\mu=N_1}^{N_2}
\Biggl(
\frac{\partial P^{I\sigma}_{m\mu}}{\partial \tau_{\alpha i}}(P^{I\sigma}_{m\mu})^{*}+
P^{I\sigma}_{m\mu}\Bigl(\frac{\partial P^{I\sigma}_{m\mu}}{\partial \tau_{\alpha i}}\Bigr)^{*}
\Biggr).
\end{eqnarray}
The derivative of the $P_{m\mu}^{I\sigma}$ can be written as
\begin{eqnarray}
\label{dP}
\frac{\partial P_{m\mu}^{I\sigma}}{\partial \tau_{\alpha i}} = \frac{\partial}{\partial \tau_{\alpha i}}\langle \phi^{I}_{m\mathbf{k}}| \hat S | \psi^{\sigma}_{\mathbf{k}\mu} \rangle =
\langle \frac{\partial \phi^{I}_{m\mathbf{k}}}{\partial \tau_{\alpha i}} | \hat S | \psi^{\sigma}_{\mathbf{k}\mu} \rangle +\nonumber\\
\sum_{st}q^{I}_{st}
\Biggl(
\langle \phi^{I}_{m\mathbf{k}} |\frac{\partial \beta^{I}_{s}}{\partial \tau_{\alpha i}} \rangle \langle\beta^{I}_{t}|\psi^{\sigma}_{\mathbf{k}\mu} \rangle + 
\langle \phi^{I}_{m\mathbf{k}} | \beta^{I}_{s} \rangle  \langle \frac{\partial \beta^{I}_{t}}{\partial \tau_{\alpha i}} |\psi^{\sigma}_{\mathbf{k}\mu} \rangle
\Biggl).
\end{eqnarray}

Calculation of atomic wavefunction derivative on atom displacement is a well-run procedure in plane-wave approaches~\cite{Himmetoglu2014}. For example, in Quantum-ESPRESSO package the derivative is calculated as:
\begin{equation}
\label{dphi}
\frac{\partial \phi^{I}_{m\mathbf{k}}}{\partial \tau_{\alpha i}}=
\delta_{I,\alpha}\frac{i}{\sqrt{V}}\sum_{\mathbf{G}}e^{-i(\mathbf{k}+\mathbf{G})\cdot\mathbf{r}}
c_{i,\alpha}(\mathbf{k}+\mathbf{G})(\mathbf{k}+\mathbf{G})_{i},
\end{equation}
where $V$ is the cell volume, $G$ is a reciprocal lattice vector and $c_{i,\alpha}$ is a Fourier coefficient of an atomic wave functions Fourier expansion.

Thus if we know the method of calculating $P^{I\sigma}_{m\mu}$ and all components in (\ref{dP}) then we can calculate the occupation matrix derivative and hence the Hubbard contribution to the atomic forces.

\section{Results and discussion}
The proposed approach was tested on one of typical objects for calculation schemes verification for correlated materials - nickel oxide. 
NiO is a charge-transfer insulator wherein the partially filled bands are formed by Ni-d and O-p
orbitals of neighboring atoms O. 
It is known that spin-polarized DFT+U calculation is able to successfully reproduce bands structure of the compound~\cite{Anisimov1991}.
Therefore we calculated an atomic forces acting on atoms at the end of self-consistent cycle within spin-polarized DFT+U approach in WF basis~\cite{Korotin2012,Korotin2013}.

For the density-functional calculations, we used the Perdew–Burke–Ernzerhof GGA exchange–correlation functional together with Vanderbilt ultrasound pseudopotentials. We used a kinetic energy cutoff of 45 Ryd (360 Ryd) for the plane-wave expansion of the electronic states (core-augmentation charge). The self-consistent calculations were performed with a (6, 6, 6) Monkhorst–Pack k-point grid. Calculations were performed for a cell containing two formula units. 
One-site effective Coulomb interaction parameter U = 8.0~eV for Ni $3d$ states was chosen \cite{Anisimov1991}.
Constructed Wannier functions had a symmetry of Ni-d and O-p states.
\begin{figure}[h!]
\centerline{\includegraphics[width=0.6\columnwidth,clip]{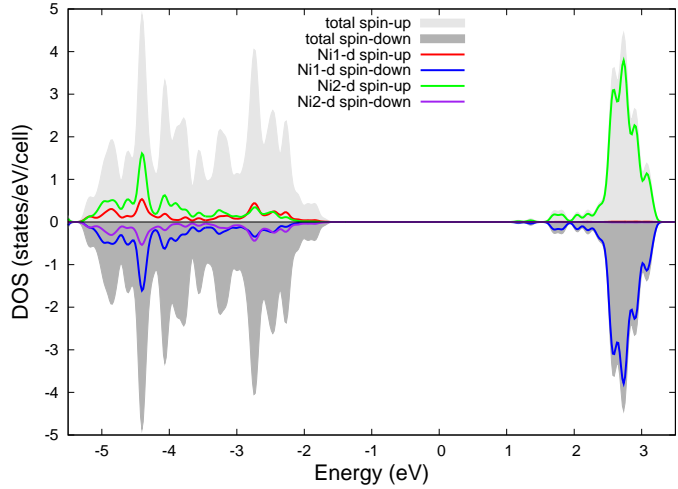}}
\caption{(color online) NiO density of states. Upper and lower panels correspond to the  of the up and down projection of the spin moment. Shaded area corresponds to the total density of states. Zero energy corresponds to Fermi level.} 
\label{fig:dos}
\end{figure}
The total and partial densities of states for Ni ions are shown on figure~\ref{fig:dos}. Obtained energy gap values equals 4~eV and one is in agreement with previous works~\cite{Anisimov1991} and photoelectron and XAS measurements~\cite{PhysRevLett.53.1951},\cite{PhysRevLett.53.2339}.
The upper Hubbard band above the Fermi level consists of Ni-d states and the lower Hubbard band is a mixture of Ni-d and O-p states due to a strong hybridization.

\begin{figure}[h!]
\centerline{\includegraphics[width=0.4\columnwidth,clip]{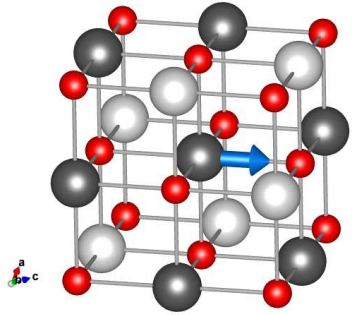}}
\caption{(color online) Schematic NiO crystal structure view with displacement of the one Ni atom (dark-gray sphere) in the $x$-direction represented by blue-color vector. Light-gray and dark-gray spheres denoted Ni atoms with different spin. Red spheres corresponding an oxygen atoms.} 
\label{fig:structure}
\end{figure}

To test the forces calculation procedure we displaced one nickel atom from its equilibrium position in the $x$-direction (the displacement direction within a cell is shown as a blue-color vector on figure~\ref{fig:structure}).
Then the Hubbard contribution to the total force were computed  using equations~(\ref{Hubforce}-\ref{dntau}).

\begin{figure}[h!]
\centerline{\includegraphics[width=0.5\columnwidth,clip]{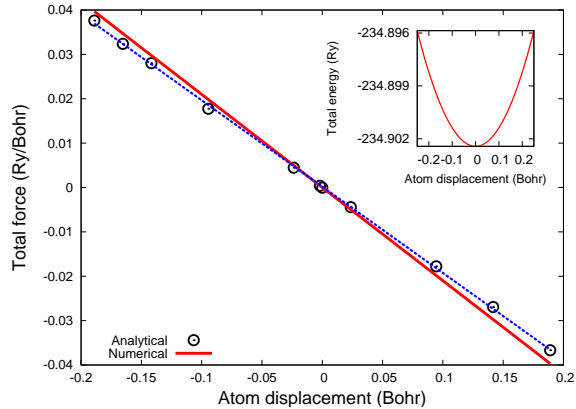}}
\caption{(color online) The dependence of total force and total energy (on inset) on displacement of one Ni atom in the $x$-direction.} 
\label{fig:total_force}
\end{figure}

The resulting total force, obtained within DFT+U in WF basis calculation, acting on the Ni atom is shown on figure~\ref{fig:total_force} with blue-color dashed line. 
The force dependence was obtained for 10 various displacement and then interpolated with a straight line. 
The "analytical" force is compared with a numerical one (red solid line on figure~\ref{fig:total_force}) obtained as a numerical derivative on total energy dependence shown on the inset of the figure.

One can see that the atomic force obtained analytically are in a good agreement with the numerical derivative. A slight difference in the lines slope could be explained as a result of neglect of some terms in equation~(\ref{dntau}).

The most clear and straightforward method for phonon frequencies calculation is the "frozen phonon" approach. The phonon frequency equals to the second derivative of the total energy over atom displacement. On the other hand it could be computed as the first derivative of the total force acting on atom. Since in both cases the derivative is computed numerically, the second approach is more accurate and it is used nowadays more intensively than the first one. From the data presented on the figure~\ref{fig:total_force} we computed the second derivative of the toal energy and the first derivative of the total force over atom displacement. The obtained values 0.21 Ry/(Bohr)$^2$ and 0.2 Ry/(Bohr)$^2$ are in a good agreement. Therefore the presented technique could be used not only for forces calculation but additionally for vibrational properties computation within the "frozen phonons" approach.





\section{Conclusion}
In the present work the approach to calculate the Hubbard term contribution to an atomic forces within Wannier functions basis into DFT+U framework with ultrasoft pseudopotential formalism is proposed.
We have performed a calculation of an atomic force acting on slightly displaced nickel atom in NiO. 
The good agreement between atomic force evaluated analytically and numerically is obtained that confirms the applicability and reliability of the proposed method.

\ack
The present work was supported by the grant of the Russian Scientific Foundation (project no. 14-22-00004).

\section*{References}
\bibliography{main}

\end{document}